\documentclass[twocolumn,amssymb, nobibnotes, showpacs, superscriptaddress, aps, prd]{revtex4}
\usepackage{amsmath}
\usepackage{epsfig}

\begin{document}
\title{Observation of a red-blue detuning asymmetry in matter-wave superradiance}
\author{L. Deng}
\affiliation{Physics Laboratory, National Institute of Standards \& Technology, Gaithersburg, Maryland 20899 USA}
\author{E.W. Hagley}
\affiliation{Physics Laboratory, National Institute of Standards \& Technology, Gaithersburg, Maryland 20899 USA}
\author{Qiang Cao}
\affiliation{Institute of Physics, Chinese Academy of Sciences, Beijing 100190, China}
\author{Xiaorui Wang}
\affiliation{Institute of Physics, Chinese Academy of Sciences, Beijing 100190, China}
\author{Xinyu Luo}
\affiliation{Institute of Physics, Chinese Academy of Sciences, Beijing 100190, China}
\author{Ruquan Wang}
\affiliation{Institute of Physics, Chinese Academy of Sciences, Beijing 100190, China}
\author{M.G. Payne}
\affiliation{Physics Laboratory, National Institute of Standards \& Technology, Gaithersburg, Maryland 20899 USA}
\author{Fan Yang}
\affiliation{School of Electronics Engineering \& Computer Science, Peking University, Beijing 100871, China}
\author{Xiaoji Zhou}
\affiliation{School of Electronics Engineering \& Computer Science, Peking University, Beijing 100871, China}
\author{Xuzong Chen}
\affiliation{School of Electronics Engineering \& Computer Science, Peking University, Beijing 100871, China}
\author{Mingsheng Zhan}
\affiliation{Center for Cold Atom Physics \& Wuhan Institute of Physics and Mathematics, Chinese Academy of Sciences, Wuhan 430071, China}
\date{\today}

\begin{abstract}
We report the first experimental observations of strong suppression of matter-wave superradiance using blue-detuned pump light and demonstrate a pump-laser detuning asymmetry in the collective atomic recoil motion. In contrast to all previous theoretical frameworks, which predict that the process should be symmetric with respect to the sign of the pump-laser detuning, we find that for condensates the symmetry is broken. With high condensate densities and red-detuned light, the familiar distinctive multi-order, matter-wave scattering pattern is clearly visible, whereas with blue-detuned light superradiance is strongly suppressed. In the limit of a dilute atomic gas, however, symmetry is restored. 
   
\end{abstract}
\pacs{03.75.-b, 42.65.-k, 42.50.Gy}

\maketitle

Matter-wave superradiance is coherent, collective atomic recoil motion that was
first reported \cite{inouye1} in a Bose-Einstein Condensate (BEC) of $^{23}$Na atoms illuminated by a single, far red-detuned, long-duration laser pulse. Since its discovery, processes such as short-pulsed, bi-directional superradiance \cite{schneble1},
Raman superradiance \cite{schneble2,kuga1}, and matter-wave
amplification \cite{inouye2,kozuma} have been observed.  
Also, many theoretical investigations \cite{moore,li,piovella,han,bonifacio,fallani,
uys,benedek,robb,ketterle} have studied this 
light/matter-wave interaction process that is of significant importance to the fields
of cold atomic physics, cold molecular physics, nonlinear optics and quantum information science.  

The widely-accepted theory \cite{inouye1} of matter-wave superradiance is based on 
spontaneous Rayleigh scattering and the buildup of a matter-wave grating enhanced by subsequent stimulated Rayleigh scattering. 
This intuitive picture, which correctly models late-stage superradiant growth when red-detuned light is used, captures many important aspects of this intriguing matter-light interaction process. However, 
the simple grating viewpoint and most rate-equation-based theories neglect propagation dynamics of the internally-generated optical field. In fact, the initial study \cite{inouye1} explicitly assumed that the optical fields traveled at the speed of light in vacuum and therefore did not affect scattering at later times. To date, no report in the literature has contradicted that statement \cite{slowwavenote}. However, we have recently shown theoretically \cite{lu1,noteref1} that the internally-generated field propagates ultra slowly and plays an important role in the genesis of superradiance with BECs. We also note that most previous theories effectively treated the BEC as a thermal gas by neglecting the extra factor of the mean-field potential seen by the scattered atoms due to the exchange term in the Hamiltonian. As we will show, this unique property of BECs profoundly impacts superradiant scattering and leads to the pump detuning asymmetry reported here.
\begin{figure}
\centering
\includegraphics[angle=90,width=3in]{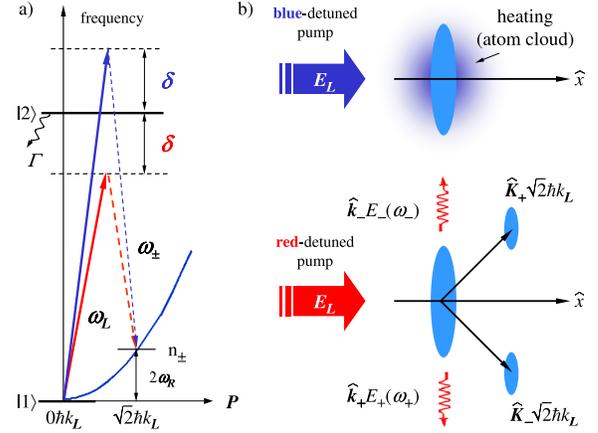} 
\caption{(a) Energy levels and laser excitation scheme 
where the one-photon detuning $\delta=\omega_L-\omega_{21}$, $\omega_{21}$ is the resonance frequency, and $\Gamma$ is the spontaneous emission rate of state $|2\rangle$. (b) Collective atomic recoil motion (matter-wave superradiance).  Upper panel: pump is blue-detuned and superradiance is strongly suppressed (only heating of the BEC).  Lower panel: pump is red-detuned and superradiance is strongly favored, resulting in a distinctive pattern of collective recoil modes. $\hat{K}_{\pm}$ ($\hat{k}_{\pm}$)
are the unit vectors for the collective atomic recoil (field) modes.}
\end{figure}

In this Letter we present the first experimental observation of a red-blue detuning asymmetry in matter-wave superradiance. We demonstrate astonishingly efficient suppression of superradiance when the pump laser is blue detuned that cannot be explained by current theoretical frameworks. However, using our new theoretical framework \cite{lu1} we propose a possible explanation for the detuning asymmetry based on an induced optical-dipole potential that results from the ultra-slow propagation velocity and gain characteristics of the generated field. 

The experimental data reported here were obtained using two $^{87}$Rb BECs created with very different experimental systems at two independent institutions. In both systems we produced an elongated BEC using standard magneto-optical trapping techniques followed by radio-frequency evaporative cooling. After formation of the BEC a pump laser of selected frequency, polarization, and duration was applied along the BEC's short axis (Fig. 1b). The magnetic trap was then switched off and absorption imaging was employed after a delay sufficient to allow spatial separation of the scattered components. For all data reported the relevant transition was $|5S_{1/2}\rangle-|5P_{3/2}\rangle$ ($|1\rangle-|2\rangle$), the ground electronic state was $F=2$, $m_F=+2$, and the detuning was measured with respect to the $F'=3$ state. We derived the pump laser from a cavity-stabilized diode laser with linear polarization perpendicular to the long axis of the BEC. The detuning asymmetry was investigated from 500 MHz $\leq |\delta|/2\pi \leq$ 4 GHz for both red and blue detunings. The blue-detuned data presented in this manuscript are entirely consistent with results obtained at other blue detunings and other pump-laser intensities. Over the range of detunings investigated, superradiance was always strongly suppressed (null result at background level) when a pure high-density BEC was illuminated with blue-detuned light. We note that the scattering efficiency for red detunings was already studied \cite{Hilliard}, and our red-detuned data are consistent with that work as well as with previous studies \cite{inouye1,schneble1,schneble2,bonifacio,fallani}.

\begin{figure}
\centering
\includegraphics[angle=90,width=3.2in]{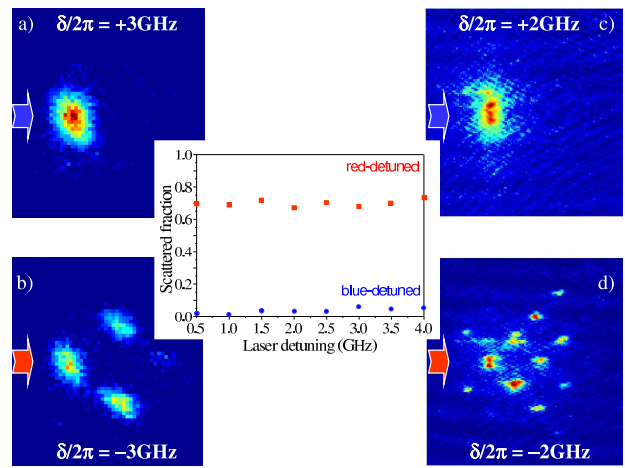} 
\caption{TOF absorption images of a BEC after application of a 200 $\mu$s pump pulse. Left panel image size is 314 $\mu$m$\times$ 336 $\mu$m, TOF = 15 ms, $5\times 10^4$ atoms, and $I_P=56$ mW/cm$^2$. The in-trap aspect ratio was about 9 to 1. (a): blue detuning, no superradiant scattering is visible. (b): red detuning, first-order superradiance is visible. Inset: detuning dependence in the case of (a) and (b). For these data the laser power was chosen to maintain a constant single-photon scattering rate, and uncertainty was typically 10 \% or less.  The scattered fraction was obtained by counting atoms in the first orders after background subtraction. We attribute the small, non-zero values with blue detunings to incomplete background subtraction and/or small residual thermal fractions. 
Right panel TOF = 20 ms, $2\times 10^5$ atoms, and $I_P=150$ mW/cm$^2$. (c): blue detuning, no superradiant scattering is visible.  (d): red detuning, higher-order scattering is evident (image size is 914 $\mu$m$\times$ 984 $\mu$m).}
\end{figure} 

The left panel of Fig. 2 shows two Time-Of-Flight (TOF) absorption images of a BEC momentum distribution after application of a pump pulse. For Fig. 2a, which shows no superradiant scattering, the laser was blue detuned by $+3$ GHz, whereas for Fig. 2b the laser was red detuned by $-3$ GHz and first-order superradiance is clearly visible. The insert in Fig. 2, which is a map of the low-power scattering efficiency for red and blue detunings, clearly shows that superradiant scattering could not be initiated with a blue-detuned pump. The right panel of Fig. 2 displays two TOF images after application of a high-power pump pulse to an elongated BEC using a different experimental apparatus. These images show that growth of higher-order momentum states is subject to a condition similar to the one that leads to suppression of first-order scattering when blue-detuned light is used. We point out that this is consistent with a sequential scattering process where higher-order growth is predicated on the growth of first-order momentum components. The above observations demonstrate the stark contrast between red- and blue-detuned pump light in the generation of collective atomic recoil motion with BECs, and raise challenges to current theoretical frameworks \cite{inouye1,schneble1,moore,li,piovella,han,bonifacio,fallani,uys,benedek,robb,ketterle} which predict the process should be symmetric with respect to detuning.

When a pump laser interacts with a BEC it first generates photons by spontaneous Rayleigh scattering, regardless of the sign of the pump-laser detuning. However, even in this early stage the BEC's structure factor \cite{sf} imposes additional constraints on the scattering process and slightly suppresses this two-photon channel to about 90$\%$ of its free-particle value. Never-the-less, these seed fields may then be amplified by coherent scattering of pump photons via the two-photon process treated in Ref. \cite{lu1}. Since the initial number of spontaneously-scattered photons per unit volume is proportional to the local density, at early times the intensity of these seed fields will directly reflect the local BEC density. However, the velocity of these growing seed fields will be inversely proportional to the local density and the field gain will be an exponential function of density. For sufficiently high spontaneous Rayleigh scattering rates the generated field will grow diabatically with respect to atomic motion, and will result in a non-negligible average optical-dipole potential $\bar{U}_{dipole}$. This induced $\bar{U}_{dipole}$ breaks the detuning symmetry of the original scattering process because for red (blue) detuned light $\bar{U}_{dipole}$ is attractive (repulsive). The important question to ask is how can $\bar{U}_{dipole}$ affect the scattering process?

It has been shown interferometrically \cite{campbell} that the energy of an atom scattered out of a BEC has an additional mean-field contribution due to the exchange term in the Hamiltonian, $E/\hbar=4\omega_{R}+\omega_{MF}$. Here $\omega_{MF}=\bar{U}_{MF}/\hbar =16\pi\hbar an_0/(7M)$ is the average mean-field shift where $a$ is the scattering length, $M$ is the atomic mass, $n_0$ is the peak condensate density, $\omega_R$ is the single-photon recoil frequency, and we have neglected the optical index of the medium because of the large detunings in this study. Clearly, the scattering is not free-particle-like because of the additional energy $\hbar \omega_{MF}$. However, with red detunings the induced $\bar{U}_{dipole}$, which is seen by both condensed and scattered atoms, will grow and eventually reach the level of the mean-field potential ($\bar{U}_{dipole}\approx -\bar{U}_{MF}$). Under this condition the net energy available to an atom scattered out of the condensate relative to the unperturbed condensate is simply $E/\hbar=4\omega_{R}$, and the scattering becomes ``free-particle-like" for all momentum transfers.  The attractive $\bar{U}_{dipole}$ can therefore be thought of as a work function for removing atoms from the BEC that is overcome by the additional factor of $\bar{U}_{MF}$ given to the scattering process by the host BEC itself. Note that only scattered atoms would experience a ``flat" potential, and that the host BEC would not be in equilibrium \cite{noteref6}. Satisfying this free-particle-like scattering condition implies that there is no extra energy left for quasi-particle excitations of the host condensate. 

This naturally brings us back to the structure factor \cite{sf} of a BEC (without $\bar{U}_{dipole}$), which goes to zero at low-momentum scatterings. If we postulate that the free-particle-scattering condition removes the constraint of the host BEC structure factor and allows low-momentum scatterings to occur, then the system would start to behave like an ultra-cold thermal gas. In this case both linear and non-linear \cite{nonlinear} processes would occur simultaneously, resulting in very efficient coherent growth of the generated field. We also point out that because $\bar{U}_{dipole}$ grows exponentially with density, it becomes more sharply peaked than the Thomas-Fermi density distribution and the resulting transverse optical-dipole force will lead to an increasing transverse velocity spread of the atoms \cite{noteref6}. We speculate that the sudden opening of efficient non-linear gain channels may facilitate triggering bosonic stimulation by creating a burst of highly monochromatic photons (atoms) scattered along (at 45$^{o}$ to) the long symmetry axis of the BEC where the transverse velocity is zero and the density is greatest. However, even without invoking non-linear gain channels the impact of the evolving structure factor on the two-photon channel may be sufficient to explain the asymmetry.

With blue-detuned light the diabatically-generated field moves the system further away from free-particle-like scattering because the growing $\bar{U}_{dipole}$ adds to $\bar{U}_{MF}$ rather than canceling it. This would cause the effective structure factor to have an increasingly larger negative impact on the two-photon channel as the optical-dipole potential grows, and would lead to gain clamping. Therefore the two-photon gain channel becomes inefficient and non-linear gain channels remain closed. In addition, the repulsive optical-dipole potential will cause a radially outward-going momentum spread, and this explains the significant expansion seen in Fig. 2c.

Although the growing 3D $\bar{U}_{dipole}$ is very difficult to model theoretically, we can estimate its importance \cite{noteref5}. 
Intuitively, photons emitted along the long axis of the BEC dominate coherent growth because of 
maximum propagation gain. From Ref. \cite{lu1}, the generated field originating at one end and propagating an effective distance $\alpha$ along the long axis results in

\begin{equation}
U(\alpha)_{dipole}\approx \hbar\left[\frac{3\lambda^2}{8\pi^2}\frac{\Gamma}{\delta}\left(\frac{N_i}{\tau_0A}\right)e^{2G\alpha}\right].\nonumber
\end{equation}
Here $\lambda$ is the generated-field wavelength, $N_i$ is the number of initial seed photons, $\tau_0$
is the pulse length of the initial seed photon burst, and $A$ is the BEC cross-section. In addition,
$G=4R\kappa_0n_0/(\gamma_B\Gamma)$ with $\kappa_0=(2\pi)^2|d|^2/(\hbar\lambda)$ where $|d|$ is the dipole transition 
matrix element, $R$ is the single-photon scattering rate, and $\gamma_B$ is the width 
of the two-photon Bragg resonance involving a pump and a generated photon. For the BEC in Ref. \cite{inouye1} when $N_i\approx 1$, $\bar{U}_{dipole}\approx-\bar{U}_{MF}$ occurs when $R\approx$ 100 Hz, in good agreement with the observed threshold scattering rate.

In the limit of thermal vapors, where there is no mean-field exchange term and the density distribution is more uniform, the process should be detuning agnostic. The scattering efficiency will be reduced because of shorter coherence times and lower density, but wave-mixing channels (both linear and non-linear) will remain open because the scattering would already be free-particle-like in nature. To test this hypothesis we applied a pump laser pulse to a BEC after adiabatically relaxing the magnetic trapping potential to lower the density. In this case wave-mixing will occur with both the condensed fraction and the uncondensed fraction that results from the expansion not being completely adiabatic. For the condensed fraction the internal-field generation will be the same as before if $R$ is increased to compensate for the lower density. However, the BEC itself will begin to look more like an ultra-cold thermal gas since $\bar{U}_{MF}$ is correspondingly reduced, bringing the initial system closer to the free-particle-scattering limit. As the system is expanded to a greater degree, less efficient generation of collective atomic recoil modes from the underlying wave-mixing processes should occur with blue detunings for both the thermal fraction and the BEC itself, and this is consistent with what we observe experimentally. For the upper images in Fig. 3, where the magnetic field was lowered to 50\% of its original value, the asymmetry is still pronounced. However for the lower images, where the magnetic field strength was lowered to 10\% of its original value (40\% uncondensed), symmetry is beginning to be restored, in agreement with our postulation.

\begin{figure}
\centering
\includegraphics[angle=90,width=3in]{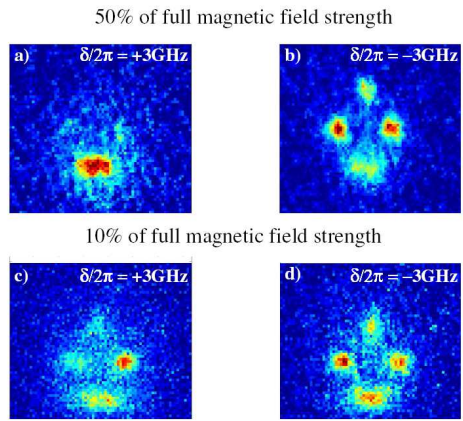}
\caption{Images showing the restoration of red-blue detuning symmetry when 
the magnetic trap is adiabatically relaxed to lower the mean-field potential. a) and b): $I_P\approx$ 60 mW/cm$^2$.
c) and d): $I_P\approx$ 110 mW/cm$^2$. In all images shown the pump duration = 200 $\mu$s, and the TOF = 12 ms. Image size: 620 $\mu$m$\times$ 612 $\mu$m, number of atoms: $2\times 10^5$. }
\end{figure}

In conclusion, we have demonstrated a red-blue detuning asymmetry in matter-wave superradiance and showed that symmetry is restored in the limit of dilute atomic vapors for which there is no additional factor of the mean-field potential to influence scattering dynamics. We also provided a plausible explanation for the symmetry breaking based on the mean-field potential of the BEC and an induced $\bar{U}_{dipole}$ to stimulate further studies. We believe that the asymmetry results from early-stage growth of a scattered optical field which causes the system to evolve toward (away from) the free-particle scattering limit with red (blue) detunings. With red detunings this results in enhanced coherent growth of an ultra-slow generated field. However with blue detunings, linear and non-linear gain channels are inhibited by the evolving structure factor, and this precludes the formation of a high contrast grating. At early times in the scattering process, where the genesis of the red-blue asymmetry occurs, the grating picture is invalid. However, at late times with red detunings, our model and previous theoretical models converge because the atomic polarization in Maxwell's equation (see Eq. 3 of Ref. \cite{lu1}) can now be viewed as a grating. We therefore believe that the origin of matter-wave superradiance is fundamentally a multi-matter-optical, wave-mixing process. The suppression of superradiance with blue detunings reported here results from the unique properties of BECs, and will therefore not occur in fermionic or uncondensed bosonic systems. Since the wave-mixing process need not invoke bosonic stimulation, collective atomic recoil motion will occur with fermions \cite{fermion}, but with a much lower efficiency. Finally we note that the widely-accepted theoretical model of matter-wave superradiance developed over the last decade is incapable of explaining our experimental results because it does not address early-stage growth of the scattering process. Since this theoretical framework provides the foundation for many important studies, its revision should be a scientific priority.

Acknowledgment: The authors acknowledge fruitful discussions with Dr. C.W. Clark, Prof. W. Ketterle, Dr. J. Bienfang, and Prof. K. Burnett. Ruquan Wang acknowledges financial support from the National Basic Research Program of China (973 project Grant No. 2006CB921206), the National High-Tech Research Program of China (863 project Grant No. 2006AA06Z104), and the National Science Foundation of China (Grant No. 10704086).


\begin{thebibliography}{13}
\bibitem{noteref1} The single-particle phase-matching condition derived in \cite{lu1} is not accurate because the photon recoil momentum in dispersive media \cite{campbell} was neglected. To first order, such static phase-matching should be satisfied for all detunings (W. Ketterle, private communications).
\bibitem{noteref6} An alternative, but consistent, viewpoint is that red and blue detunings chirp the BEC phase in different ways, and modify the light-scattering properties of the superfluid that now finds itself under compression or tension.
\bibitem{noteref5} We believe that when the Gross-Pitasvskii and Maxwell equations are solved self-consistently, a generalized dynamic phase-matching relation including the polarization force/potential and the BEC structure factor will verify our postulation about the underlying mechanism.
\end{thebibliography}
\end{document}